\documentclass[11pt]{article}
\usepackage{listings}
\usepackage{graphicx,url}
\usepackage[utf8]{inputenc}
\usepackage{amsthm}
\usepackage{newtxtext}    

\usepackage{amsmath,amsthm,amssymb,epsfig,graphics,psfrag,latexsym,amsmath,listings,graphicx,url,tabularx,gensymb,stmaryrd}
\usepackage{algorithm}
\usepackage[utf8]{inputenc}
\usepackage[noend]{algpseudocode}
\usepackage{multicol}
\usepackage{bbm}
\usepackage{amsrefs}
\usepackage{color}
\graphicspath{ {Figures/} }
\usepackage{geometry}
\usepackage[colorinlistoftodos,textsize=tiny]{todonotes}
\newcommand{\Comments}{1}
\newcommand{\mynote}[2]{\ifnum\Comments=1\textcolor{#1}{#2}\fi}
\newcommand{\mytodo}[2]{\ifnum\Comments=1%
  \todo[linecolor=#1!80!black,backgroundcolor=#1,bordercolor=#1!80!black]{#2}\fi}

\def\argmax{{\hbox{argmax}}}

\title{Metric Dimension}

\author{
  Richard C. Tillquist\\
  \texttt{richard.tillquist@colorado.edu}
  \and
  Rafael M. Frongillo  \\
  \texttt{raf@colorado.edu}
  \and
  Manuel E. Lladser \\
  \texttt{manuel.lladser@colorado.edu}
}

\begin{document}

\date{}

\maketitle

The \textbf{metric dimension} of a graph is the smallest number of vertices from which the vector of distances to every vertex in the graph is unique. It may be regarded as a generalization of the concept of trilateration in the two-dimensional real plane, the idea underpinning the Global Positioning System (GPS).

\section{Definition}

\begin{figure}[h]
\centering
\includegraphics[scale=0.65]{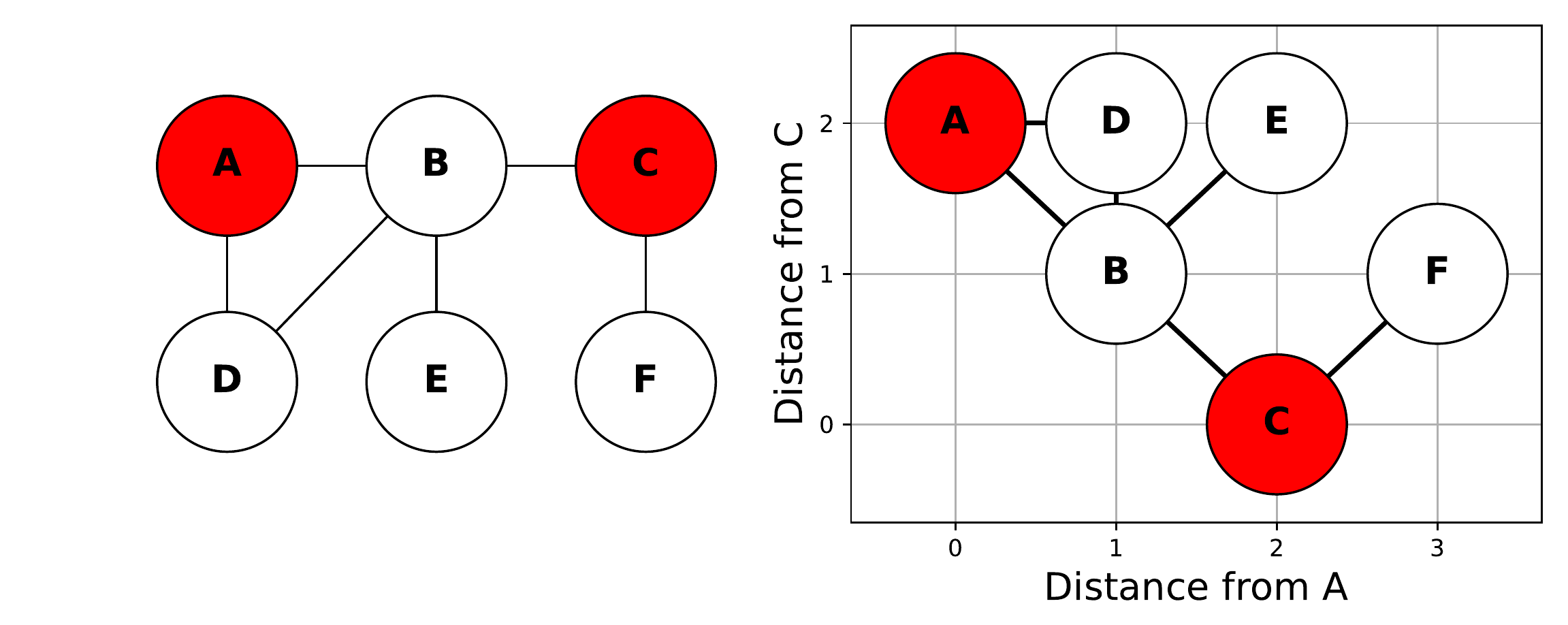}
\caption[LoF entry]{A graph with metric dimension 2 and minimal resolving set $R=\{A, C\}$. Based on this set, $\Phi_R(A) = (0,2)$, $\Phi_R(B) = (1,1)$, $\Phi_R(C) = (2,0)$, $\Phi_R(D) = (1,2)$, $\Phi_R(E) = (2,2)$, and $\Phi_R(F) = (3,1)$. This corresponds to the embedding of the graph in $\mathbb{R}^2$ on the right.}
\label{fig:metric_dimension_ex}
\end{figure}

Let $G$ be a graph with vertex set $V$ and edge set $E$, and let $d(u,v)$ denote the shortest path or geodesic distance between two vertices $u,v \in V$. $G$ is not forced to be simple (though all examples in this article are) and may contain weighted edges, multi-edges, or self loops. A set $R \subseteq V$ is called \textbf{resolving} if for all $u,v \in V$ with $u \neq v$ there is at least one $r \in R$ such that $d(u,r) \neq d(v,r)$. In this case $r$ is said to \emph{resolve} or \emph{distinguish} $u$ and $v$. By definition, if an ordering on the vertices of $R=\{r_1,\dots,r_n\}$ is given, any $u \in V$ may be uniquely represented by the vector $\Phi_R(u) := (d(u,r_1), \dots, d(u,r_n))$ (see Figure~\ref{fig:metric_dimension_ex}). The \textbf{metric dimension} of $G$, denoted $\beta(G)$, is the smallest size of resolving sets on $G$; formally, $\beta(G) = \min\{|R| : R \subseteq V, \; R \text{ is resolving}\}$. If $R$ is a resolving set on $G$ and $|R| = \beta(G)$, $R$ is called a \textbf{minimal resolving set} of $G$, also called a \emph{basis set}, or \emph{reference set}~\cite{harary1976metric,slater1975leaves}.

Intuitively, this concept is closely related to that employed by the Global Positioning System (GPS), called \textbf{trilateration}, where the location of any object on Earth can be determined by its distances to three satellites in orbit. More generally, given a point $x \in \mathbb{R}^2$, we may partition the space into equivalence classes of points with equal Euclidean distance to $x$, where $y,z \in \mathbb{R}^2$ belong to the same class if and only if $d(y,x) = d(z,x)$ (these classes form circles centered at $x$). A set of points $R \subset \mathbb{R}^2$ may be used to partition the space in a similar way. Now $y$ and $z$ belong to the same class if and only if $d(y, r) = d(z, r)$ for all $r \in R$. When $R$ contains a subset of three affinely independent points, every point in $\mathbb{R}^2$ belongs to its own equivalence class and $R$ may be said to resolve the plane.

\section{Brute Force Calculation}

Given an arbitrary graph $G=(V,E)$, the brute force method for determining $\beta(G)$ requires that every subset of $(\beta(G)-1)$ vertices be established as non-resolving and that at least one resolving set of size $\beta(G)$ be found. Since $\beta(G) \leq |V|-1$~\cite{chartrand2000resolvability}, starting with sets of size one, $\sum_{k=1}^{|V|-2} \binom{|V|}{k} = O(2^{|V|})$ subsets must be examined in the worst case. In order to determine whether or not $R \subseteq V$ is resolving, every pair of vertices $u,v \in V$ must be compared across $|R|$ distances. This requires $O(|R||V|^2)$ time, bringing the total time necessary to find $\beta(G)$ to $|V|^2\sum_{k=1}^{|V|-2}\binom{|V|}{k}k=O(|V|^32^{|V|})$.

The above assumes that all pairwise distances between nodes in $G$ have been precomputed. There are a host of algorithms for finding shortest path distances in graphs. When $G$ is directed and may have positive or negative edge weights, the Floyd-Warshall algorithm and Johnson's algorithm are among the most popular techniques. These have asymptotic run times $O(|V|^3)$~\cite{floyd1962algorithm} and $O(|V||E|+|V|^2\log|V|)$~\cite{johnson1977efficient}, respectively. An algorithm based on a component hierarchy~\cite{thorup1999undirected} can solve this problem in $O(|V||E|+|V|^2\log\log|V|)$ time~\cite{pettie2004new}. When $G$ is undirected and edge weights are guaranteed to take integer values, a similar approach can be used to determine all shortest path lengths in $O(|V||E|)$ time~\cite{thorup1999undirected}.

\section{Complexity and Approximation Algorithms}

The brute force approach to computing $\beta(G)$ is intractable even for small graphs. In fact, this problem is NP-hard and the associated decision problem, determining whether the metric dimension of a graph is less than a specified integer, has been shown to be NP-complete via reduction from 3-SAT~\cite{khuller1996landmarks} and 3-dimensional matching~\cite{gary1979computers}. As a result, a number of algorithms for estimating metric dimension exist. Methods employing genetic algorithms~\cite{kratica2009computing} and a variable neighborhood search~\cite{mladenovic2012variable} can find small resolving sets but do not provide approximation guarantees which bound how far from $\beta(G)$ the result may be. The \textbf{Information Content Heuristic (ICH)}, on the other hand, ensures an approximation ratio of $1 + (1 + o(1))\cdot\ln(|V|)$, the best possible ratio for metric dimension~\cite{hauptmann2012approximation}.

A brief description of the ICH algorithm follows. Let $u_R= \Phi_R(u)$ be the vector of distances from $u \in V$ to the elements of $R \subseteq V$. Let $S_R = \{u_R | u \in V\}$ be the set of all such vectors for a given graph and 
$B_R = [ u_R | u \in V ]$ be the bag or multiset associated with $S_R$. The ICH algorithm takes an information theoretic perspective, using $H(B_R)$, the discrete entropy over the multiset of vertex representations on $V$ imposed by $R$, to measure how far $R$ is from being resolving. Notice $H(B_R)$ is maximized precisely when $R$ is a resolving set, i.e. $|S_R| = |V|$ so that every vertex has a unique representation. At its core, the ICH algorithm is a greedy search for an $R$ achieving this maximum value, $H(B_R) = \log|V|$. Starting with $R_0 = \emptyset$, $R_i$ is built recursively by finding $v^* = \argmax_{v \in V \setminus R_{i-1}} H(R_{i-1} \cup \{v\})$ and setting $R_i = R_{i-1} \cup \{v^*\}$.

With a run time complexity of $O(|V|^3)$, ICH is only practical for small and medium-sized graphs. Nevertheless, using parallel computing, it is possible to reduce the run time of the ICH algorithm further.

\section{Metric Dimension of Specific Graph Families}

While determining the exact metric dimension of an arbitrary graph is a computationally difficult problem, efficient algorithms, exact formulae, and useful bounds have been established for a variety of graphs. This section presents descriptions of the metric dimension of several common families of graphs. For a list of results related to the join and cartesian product of graphs, see~\cite{caceres2005metric}. 

\vspace{5mm}

\begin{figure}[h]
\centering
\includegraphics[scale=0.65]{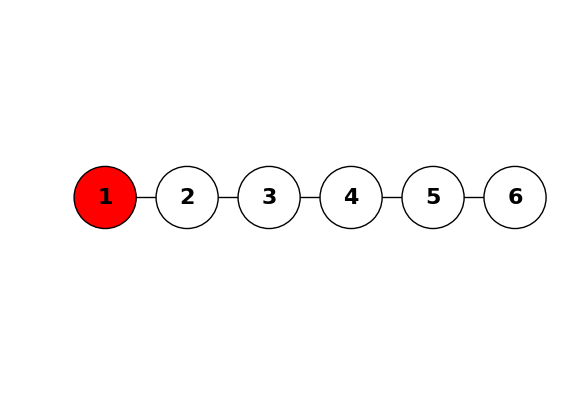}
\caption[LoF entry]{The path graph of size 6, $P_6$. $R = \{1\}$ is a minimal resolving set of this graph. In general, any set of the form $\{v\}$, with $v$ a node of degree 1 in $P_n$, is a minimal resolving set on $P_n$.}
\label{fig:path}
\end{figure}

\begin{figure}[h]
\centering
\includegraphics[scale=0.65]{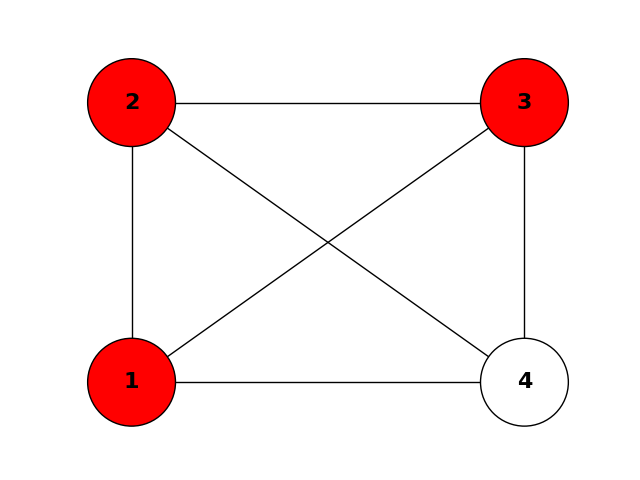}
\caption[LoF entry]{The complete graph of size 4, $K_4$. $R = \{1, 2, 3\}$ is a minimal resolving set of this graph. In general, any set of nodes of cardinality $(n-1)$ is a minimal resolving set of $K_n$.}
\label{fig:complete}
\end{figure}

\noindent \textbf{Fully Characterized Graphs:} Graphs on $n$ vertices with a metric dimension of 1, $(n-1)$, and $(n-2)$ have been fully characterized~\cite{chartrand2000resolvability}.
The first two cases are simple to describe:

\begin{itemize}
    \item The metric dimension of a graph is 1 if and only if the graph is a path (see Figure~\ref{fig:path}).
    \item The metric dimension of a graph with $n$ nodes is $(n-1)$ if and only if the graph is the complete graph on $n$ nodes (see Figure~\ref{fig:complete}).
\end{itemize}

For the third case, let us introduce notation, following~\cite{chartrand2000resolvability}.
Let $G \cup H$ be the \textbf{disjoint union} of two graphs $G$ and $H$, i.e. if $G=(V_1,E_1)$ and $H=(V_2,E_2)$, $G \cup H=(V_1 \sqcup V_2,E_1 \sqcup E_2)$, where $\sqcup$ denotes disjoint set union~\cite{WikiDisUni}. Further, let $G + H$ be the graph $G \cup H$ with additional edges joining every node in $G$ with every node in $H$. Finally, define $K_n$ to be the complete graph on $n$ nodes, $\overline{K_n}$ to be the graph with $n$ nodes and no edges, and $K_{n,m}$ to be the complete bipartite graph with partitions of size $n$ and $m$. Then the metric dimension of a graph with $n$ nodes is $(n-2)$ if and only if the graph is one of the following:

\begin{itemize}
    \item $K_{s,t}$ with $s,t \geq 1$, and $n=s+t$.
    \item $K_s + \overline{K_t}$ with $s \geq 1$, $t \geq 2$, and $n=s+t$.
    \item $K_s + (K_1 \cup K_t)$ with $s, t \geq 1$, and $n=s+t+1$.
\end{itemize}

\vspace{5mm}

\begin{figure}[h]
\centering
\includegraphics[scale=0.65]{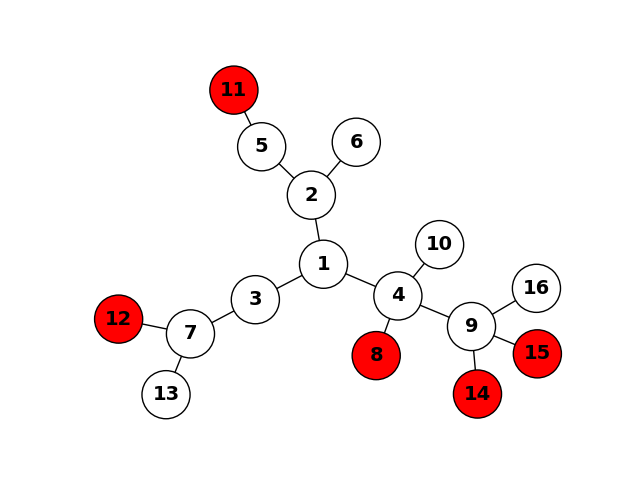}
\caption[LoF entry]{A tree of size 16. The vertices 2, 4, 7, and 9 are exterior major vertices and 6, 8, 10, 11, 12, 13, 14, 15, and 16 are leaves. Note that node 1 is not an exterior major vertex as every path from this vertex to a leaf includes at least one other vertex of degree greater than two. $R = \{8,11,12,14,15\}$ is a resolving set of minimum size.}
\label{fig:tree}
\end{figure}

\noindent \textbf{Trees:} The introduction of metric dimension in the mid 1970s also brought a characterization of the metric dimension of trees, via a simple formula~\cite{harary1976metric,slater1975leaves}.
Let $T$ be a tree that is not a path and define $\ell(T)$ to be the number of \textbf{leaves} (nodes of degree 1) in $T$. Further, define $\sigma(T)$ as the number of \textbf{exterior major vertices} in $T$, that is vertices with degree at least 3 which are also connected to at least one leaf by a path of vertices of degree 2. Then the metric dimension of $T$ is $\beta(T) = \ell(T) - \sigma(T)$. A resolving set of this size may be constructed by taking the set of all leaves and removing exactly one element associated with each exterior major vertex~\cite{chartrand2000resolvability} (see Figure~\ref{fig:tree}). This construction may be carried out using a modified depth first search in $O(|V|+|E|)$ time.

\vspace{5mm}

\noindent \textbf{Hamming Graphs:} For positive integers $k$ and $a$, the Hamming graph $H_{k,a}$ consists of $a^k$ vertices each labeled with a unique string of length $k$ using an alphabet of size $a$. Two vertices in $H_{k,a}$ are adjacent when their labels differ in exactly one position; thus, the shortest path distance $d(u,v)$ is the total number of mismatches between the labels of $u$ and $v$ (i.e. the \textbf{Hamming distance} between $u$ and $v$). While determining $\beta(H_{k,a})$ exactly is difficult, it has been shown that, in general, $\beta(H_{a,k}) \leq \beta(H_{a, k+1}) \leq \beta(H_{a,k}) + \lfloor \frac{a}{2} \rfloor$. Furthermore, given a resolving set on $H_{k, a}$ of size $s$ it is possible to efficiently construct a resolving set on $H_{k+1,a}$ of size $s + \lfloor \frac{a}{2} \rfloor$~\cite{TilLla18}. This implies that $\beta(H_{k,a})$ grows at most linearly with $k$ and allows small resolving sets to be generated despite how quickly Hamming graphs grow in size with increasing $k$.

Connections between coin weighing problems and $Q_k=H_{k,2}$, or hypercubes, lead to the asymptotic formula $\lim_{k \rightarrow \infty} \beta(Q_k) \frac{\log(k)}{k} = 2$~\cite{erdos1963two,lindstrom1964combinatory}. Even with a binary alphabet, $\beta(Q_k)$ is known exactly only up to $k=10$ (see Table~\ref{tab:hypercube_betas}).

\begin{table}[h!]
  \centering
  \begin{tabular}{|c|c|c|c|c|c|c|c|c|c|c|c|c|c|c|c|c|c|}
    \hline
    k & 1 & 2 & 3 & 4 & 5 & 6 & 7 & 8 & 9 & 10 & \emph{11} & \emph{12} & \emph{13} & \emph{14} & \emph{15} & \emph{16} & \emph{17} \\ \hline
    $\beta(Q_k)$ & 1 & 2 & 3 & 4 & 4 & 5 & 6 & 6 & 7 & 7 & \emph{8} & \emph{8} & \emph{8} & \emph{9} & \emph{9} & \emph{10} & \emph{10} \\ \hline
  \end{tabular}
  \caption{Exact values of $\beta(Q_k)$ for $1 \leq k \leq 10$, and upper bounds for $11 \leq k \leq 17$~\cite{mladenovic2012variable}.}
  \label{tab:hypercube_betas}
\end{table}

The Hamming graph $H_{k,a}$ may also be thought of as the \textbf{cartesian product} of $k$ complete graphs of size $a$. That is, $H_{k,a} = K_a \square K_a \square \dots \square K_a$, with $k$ copies of $K_a$. In general, $G \square H$, the cartesian product of $G=(V_1,E_1)$ and $H=(V_2,E_2)$, has vertex set $V = \{(u,v) | u \in V_1, v \in V_2\}$ and edge set $E$ defined as follows: $\{(u,v), (u',v')\} \in E$ if and only if $u=u'$ and $\{v,v'\} \in E_2$, or $v=v'$ and $\{u,u'\} \in E_1$. Working from this perspective, it has been shown that $\beta(H_{2,a}) = \lfloor \frac{2}{3}(2a-1) \rfloor$~\cite{caceres2007metric}.

\vspace{5mm}

\noindent \textbf{Random Graphs:} In a study related to the graph isomorphism problem, it was shown that the set of $\lceil \frac{3\ln(n)}{\ln(2)} \rceil$ high degree vertices in a graph of size $n$ can be used to differentiate two random graphs with high probability~\cite{babai1980random}. Indeed, this set of nodes is highly likely to resolve the Erd\"os-R\'enyi random graph $G_{n,1/2}$. This bound has been generalized to encompass arbitrary values of $p$ so that, with high probability, $\beta(G_{n,p}) \leq \frac{-3\ln(n)}{\ln(p^2+(1-p)^2)}$ as $n$ goes to infinity and any set of nodes of this size resolves the graph with high probability~\cite{TilLla19sbm}. Focusing closely on different regimes of $p$ as a function of the graph size, much more precise bounds on $\beta(G_{n,p})$ have been established~\cite{bollobas2012metric}. 

Closely related to Erd\"os-R\'enyi random graphs are graphs generated via the Stochastic Block Model (SBM). This model groups a set of $n$ vertices into communities defined by a partition $C$ of $\{1, \dots, n\}$. Adjacency probabilities for vertices in different communities are defined by a matrix $P$. By focusing on this adjacency information, general bounds on $G \sim SBM(n;C,P)$ have been established as have several efficient algorithms for finding small resolving sets on $G$ when $n$ is large enough to render the ICH algorithm impractical~\cite{TilLla19sbm}.

Random trees and forests have also been investigated with respect to metric dimension~\cite{mitsche2015limiting}. The exact formula and polynomial time algorithm for finding minimal resolving sets on trees allow the limiting distribution of $\beta(T_n)$, the metric dimension of a tree or forest chosen uniformly at random from all trees or forests of size $n$, to be determined precisely. In particular, 
\begin{equation*}
    \frac{\beta(T_n) - \mu n (1+o(1))}{\sqrt{\sigma^2 n(1+o(1))}} \rightarrow N(0,1),
\end{equation*}
where the convergence is in distribution as $n \rightarrow \infty$, and $\mu \simeq 0.14076941$ and $\sigma^2 \simeq 0.063748151$.

\section{Applications}

Despite the fact that finding minimal resolving sets of general graphs is computationally difficult, the ability to uniquely identify all vertices in a graph based on distances has proven to be quite useful. Applications regarding chemical structure~\cite{chartrand2000resolvability} and robot navigation~\cite{khuller1996landmarks} have served as inspiration for the theoretical study of metric dimension. Deep connections between the metric dimension of Hamming graphs and a complete understanding and analysis of the game Mastermind~\cite{chvatal1983mastermind} and various coin weighing problems~\cite{erdos1963two,lindstrom1964combinatory} have also been established. Resolving sets have proven valuable in a number of other applications as well.

\vspace{5mm}

\noindent \textbf{Source Localization:} Resolving sets are a natural tool to identify the source of a diffusion across a network. For instance, the ability to determine where a disease began as it spreads across a community has the potential to be valuable in a variety of contexts. If the time at which the spread began is known, and inter-node distances are deterministic and known, resolving sets give a direct solution. In more realistic settings, however, the notion of resolvability must be augmented to take into account an unknown start time and random transmission delays between nodes. The former may be addressed using doubly resolving sets. Whereas for every pair of different nodes $u,v \in V$ a resolving set $R \subseteq V$ need only contain a single element $r \in R$ such that $d(u,r) \neq d(v,r)$, a doubly resolving set $D \subseteq V$ must have nodes $r_1, r_2 \in D$ such that $d(u,r_1)-d(u,r_2) \neq d(v,r_1)-d(v,r_2)$. Successfully identifying the source of a spread is highly dependent on the variance associated with random inter-node distances~\cite{spinelli2016observer}.

\vspace{5mm}

\noindent \textbf{Representing Genetic Sequences:} Many machine learning algorithms assume numeric vectors as input. In contrast, sequences of nucleotides or amino acids from biological applications are symbolic in nature; as such, they must be transformed before they can be analyzed using machine learning techniques.
One such transformation is an embedding based on resolving sets, which can be used to efficiently generate concise feature vectors for large sequences.
In this approach, all possible sequences of length $k$ are encoded as nodes in a Hamming graph $H_{k,a}$, where $a$ is a reference alphabet size; given a resolving set $R$ of $H_{k,a}$, each vertex $v$ maps to the point $\Phi_R(v) \in \mathbb{R}^{|R|}$ (see Figure~\ref{fig:metric_dimension_ex}).
For example, consider $H_{8,20}$, the Hamming graph used to represent amino acid sequences of length $k=8$. This graph has approximately $25.6$ billion vertices and $1.9$ trillion edges, making many state-of-the-art graph embedding methods like multidimensional scaling~\cite{Krz00} and Node2Vec~\cite{grover2016node2vec} impractical.
On the other hand, a resolving set of size $82$ is known for this graph, which was constructed by augmenting a resolving set for $H_{3,20}$ using bounds described in Section 4~\cite{TilLla18}.
This resolving set gives rise to an embedding into $\mathbb{R}^{82}$, whereas traditional techniques used to embed biological sequences, like binary vectors, require almost twice as many dimensions.



\section{Acknowledgements}

This article was partially funded by NSF IIS grant 1836914.

\bibliographystyle{amsrefs}

\end{document}